# Index-antiguiding in narrow-ridge GaN-based laser diodes investigated by measurements of the current-dependent gain and index spectra and by self-consistent simulation

L. Redaelli, H. Wenzel, J. Piprek, T. Weig, S. Einfeldt, M. Martens, G. Lükens, U. T. Schwarz, and M. Kneissl

***Abstract***: The threshold current density of narrow (1.5 µm) ridge-waveguide InGaN multi-quantum-well laser diodes, as well as the shape of their lateral far-field patterns, strongly depend on the etch depth of the ridge waveguide. Both effects can be attributed to strong index-antiguiding. A value of the antiguiding factor R = 10 is experimentally determined near threshold by measurements of the current-dependent gain and refractive index spectra. The device performances are simulated self-consistently solving the Schrödinger-Poisson equations and the equations for charge transport and waveguiding. Assuming a carrier-induced index change which matches the experimentally determined antiguiding factor, both the measured high threshold current and the shape of the far-field pattern of lasers with shallow ridges can be reproduced theoretically.

## I. INTRODUCTION

TODAY single-lateral-mode laser diodes emitting in the blue and violet regions of the spectrum find application in several fields, including optical data storage and spectroscopy, and are entering the market of laser projection [1,2]. In order to obtain stable emission by just one single lateral mode, it is essential that the shape and height of the ridge waveguide (RW) are chosen carefully [3,4]. In the (Al,In,Ga)N material system ridge widths of 2 µm or less are typically necessary to suppress the oscillation of higher-order modes [5].

It has been recently shown [6] that a small difference in the ridge etch depth of otherwise identical blue-emitting laser diodes can cause a large difference in the threshold current density. In order to explain such behavior, a simplified model based on strong antiguiding effects has been proposed [6]. By assuming a gain-independent antiguiding factor of 10 and step-like lateral distributions of the gain and the carrier-induced refractive index change in the multi-quantum well (MQW) structure, the large difference in threshold current could be reproduced. If the ridge is not deep enough, the carrier-induced index change can compensate the built-in refractive index step, causing radiation losses and an increase in the threshold current. The model was capable of reproducing the peculiar shape of the lateral far-field patterns of shallow-ridge devices where small side lobes appear on both sides of the main Gaussian peak. The side-lobes were attributed to the tilted phase-front of the mode due to antiguiding [7]. However, in the previous work, [6] the assumed high value (R = 10) of the antiguiding factor was not validated by measurements. Moreover, the step-like gain and refractive index model strongly simplified the real profiles.

In the present paper, after briefly introducing the devices under study, the antiguiding factor is determined by measurements of the current-dependent gain and index spectra. Realistic 2D electro-optical simulations of the laser diodes are then performed, and the results are compared to the experiment.

This work has been partly funded by the German ministry of research (BMBF) through the regional growth core Berlin WideBaSe under contracts 03WKBT03B and 03WKBT03C.

L. Redaelli was with the Ferdinand-Braun-Institut, Leibniz-Institut für Höchstfrequenztechnik, Gustav Kirchhoff Str. 4, 12489 Berlin, Germany. He is now with the French Alternative Energies and Atomic Energy Commission, 17 rue des Martyrs, 38000 Grenoble, France. (e-mail: luca.redaelli@cea.fr)

H. Wenzel and S. Einfeldt are with the Ferdinand-Braun-Institut, Leibniz-Institut für Höchstfrequenztechnik, Gustav Kirchhoff Str. 4, 12489 Berlin, Germany. (e-mails: hans.wenzel@fbh-berlin.de, sven.einfeldt@fbh-berlin.de)

J. Piprek is with the NUSOD Institute LLC, P.O. Box 7204, Newark, DE 19714, USA. (e-mail: piprek@nusod.org)

T. Weig is with the Fraunhofer Institute for Applied Solid State Physics IAF, Tullastr. 72, 79108 Freiburg, Germany. (e-mail: thomas.weig@iaf.fraunhofer.de)

M. Martens is with the Institute for Solid State Physics, Technische Universität Berlin, Hardenbergstr. 36, 10623 Berlin, Germany (e-mail: martens@mailbox.tu-berlin.de).

G. Lükens was with the Fraunhofer Institute for Applied Solid State Physics IAF, Tullastr. 72, 79108 Freiburg, Germany. He is now with RTWH Aachen, Sommerfeldstraße 24, 52074 Aachen, Germany (e-mail: luekens@gan.rwth-aachen.de).

U. T. Schwarz is with the Fraunhofer Institute for Applied Solid State Physics IAF, Tullastr. 72, 79108 Freiburg, Germany and the Department of Microsystems Engineering (IMTEK), Univ. of Freiburg, Georges-Köhler-Allee 103, 79110 Freiburg, Germany (e-mail: ulrich.schwarz@iaf.fraunhofer.de;).

M. Kneissl is with the Ferdinand-Braun-Institut, Leibniz-Institut für Höchstfrequenztechnik, Gustav Kirchhoff Str. 4, 12489 Berlin, Germany and the Institute for Solid State Physics, Technische Universität Berlin, Hardenbergstr. 36, 10623 Berlin, Germany (e-mail: kneissl@physik.tu-berlin.de).

## II. Experimental Results

### A. Device Fabrication and Characterization

The investigated InGaN multiple-quantum-well (MQW) RW laser diodes, emitting at about 440 nm, had a ridge width and a resonator length of 1.5 μm and 600 μm, respectively. Their facets were uncoated. The heterostructure, shown schematically in Fig. 1, consisted of an InGaN MQW active region sandwiched between step-graded InGaN/GaN waveguides and AlGaN cladding layers. The high-Al-content p-AlGaN electron blocking layer (EBL) was placed in the p-waveguide at the InGaN/GaN interface, 70 nm above the MQW. The depth of the active region, i.e. the vertical distance from the surface to the top quantum well (QW) was about 650 nm. Two sets of laser diodes were fabricated, which only differed in the etch depth of the ridge waveguide. The etch depths were 620 nm in the "deep-ridge" laser diodes and 450 nm in the "shallow-ridge" laser diodes. These values correspond to residual layer thicknesses above the MQW of $d_{RES,deep}$ = 30 nm and $d_{RES,shallow}$ = 200 nm, and built-in effective-index steps $\Delta n_{eff,deep}$ = 0.026 and $\Delta n_{eff,shallow}$ = 0.004 for the deep-ridge and the shallow-ridge devices, respectively. The thickness of each one of the four InGaN QWs was less than 4 nm, and the vertical confinement factor of the mode was $\Gamma$ = 0.038 assuming a laterally infinite layer structure and neglecting the antiguiding effect. The epitaxial structure, including the real part of the refractive indices used in the simulations, is summarized in Table I. Further details on the structure and the fabrication process have been reported elsewhere [6],[8].

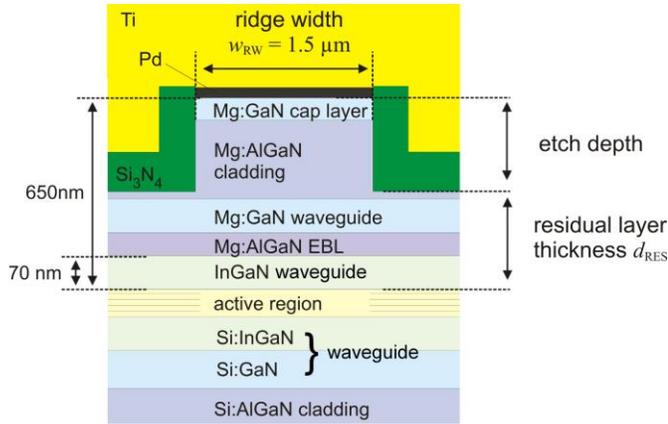

Fig. 1. Schematic cross-section drawing of the fabricated devices.

A large number of devices with uncoated facets were measured in pulsed operation (300 ns, 1 kHz), to avoid device heating. The threshold current ($I_{th}$) was between 90 mA and 110 mA for deep-ridge devices and between 220 mA and 330 mA for shallow-ridge devices. The scattering of the

TABLE I
EPITAXIAL STRUCTURE

| Layer | Material | Thickness (nm) | Index (real) |
|---|---|---|---|
| Contact layer | GaN:Mg | 40 | 2.4811 |
| Cladding | AlGaN:Mg | 470 | 2.4513 |
| Opt. conf. layer | GaN:Mg | 50 | 2.4811 |
| EBL | AlGaN:Mg | 20 | 2.4025 |
| Spacer | InGaN:Mg | 70 | 2.5019 |
| MQW | InGaN | 44.5 | 2.5800 |
| Spacer | InGaN:Si | 70 | 2.5019 |
| Opt. conf. layer | GaN:Si | 70 | 2.4811 |
| Cladding | AlGaN:Si | 1100 | 2.4513 |
| Buffer | GaN:Si | 1000 | 2.4811 |

The refractive indices have been calculated according to the model used in Ref [9]

experimental values is attributed to spatial non-uniformities in the epitaxy and in the fabrication process. In Fig. 2, the L–I characteristics and lateral (slow-axis) far-field patterns of representative devices are shown.

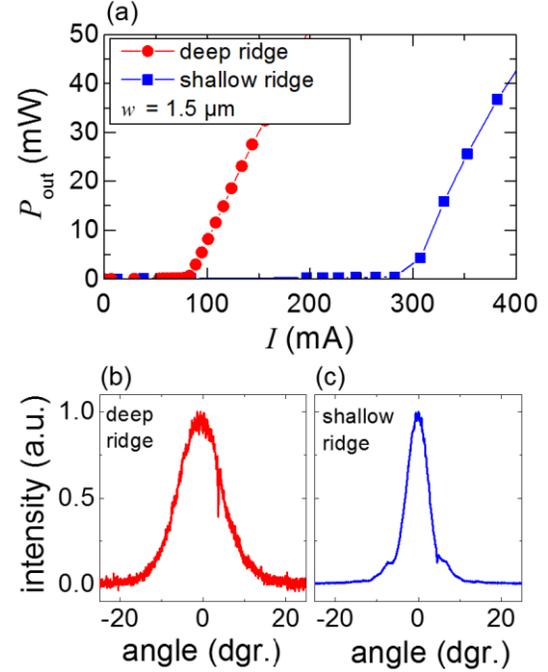

Fig. 2. Typical L−I characteristics (a) and lateral far-field patterns [(b) and (c)] of deep-ridge and shallow-ridge devices. The facets are uncoated, and the emission measured from only one facet.

### B. Determination of the antiguiding factor

A common figure of merit for the strength of antiguiding in a laser diode is the antiguiding factor R, defined as:

$$R = -\frac{4\pi}{\lambda} \frac{dn_r / dN}{dg / dN}, \quad (1)$$

where $n_r$ is the refractive index, g is the material gain, N is the excess carrier density, and $\lambda$ is the wavelength. In the experiment, the modal differential gain and modal refractive index change are measured as function of current:

$$R_{\exp} = -\frac{4\pi}{\lambda} \frac{dn_{MOD}/dI}{dg_{MOD}/dI}, \quad (2)$$

where $n_{MOD}$ is the modal index, $g_{MOD}$ the modal gain, and $I$ the current. The antiguiding factor $R$ and the experimentally accessible value $R_{\exp}$ are equal as long as the optical confinement factor $\Gamma \approx g_{MOD}/g$ is independent of the carrier density. This condition is assumed to be satisfied in the deep-ridge device. In the present section, the value of $R_{\exp}$ will be determined experimentally according to (2). In Section III the value of $R$ necessary to theoretically reproduce the measurement data will be calculated and compared to $R_{\exp}$.

The wavelength-dependent differential modal gain, $dg_{MOD}(\lambda)/dI$, was determined for the deep-ridge devices by the Hakki-Paoli method [10],[11]. From the modulation amplitude of the spectra of amplified spontaneous emission below threshold, the modal gain spectra $g_{MOD}(\lambda)$ were obtained, as shown in Fig. 3(a). The measurement was performed in the current range 30 mA to 90 mA (in this specific device $I_{th}$ = 90 mA), at intervals of 5 mA. The peak modal gain increases almost linearly with current, which translates into $dg_{MOD}/dI$ = 0.7 cm$^{-1}$mA$^{-1}$.

The differential modal index, $dn_{MOD}/dI$, was derived from the spectral shift of the longitudinal modes between two consecutive current steps as described by Scheibenzuber et al. [12]. The obtained values are assumed to be the differential index change in the center of each 5 mA interval. Self-heating and the temperature-induced change of the modal index were avoided by keeping the waveguide temperature constant after determination of the thermal resistance [12]. The obtained $dn_{MOD}/dI$ spectra are shown in Fig. 3(b).

From this data the value of the antiguiding factor at the peak wavelength of the modal gain spectra was determined, and is plotted in Fig. 3(c) as a function of current. The antiguiding factor is more than 50 at 30 mA and approaches a value of about 10 slightly below threshold. The value $R_{\exp} \approx 10$ for the antiguiding factor at threshold is very large in comparison to other material systems [13],[14], but it is not untypical for InGaN MQW laser diodes [15].

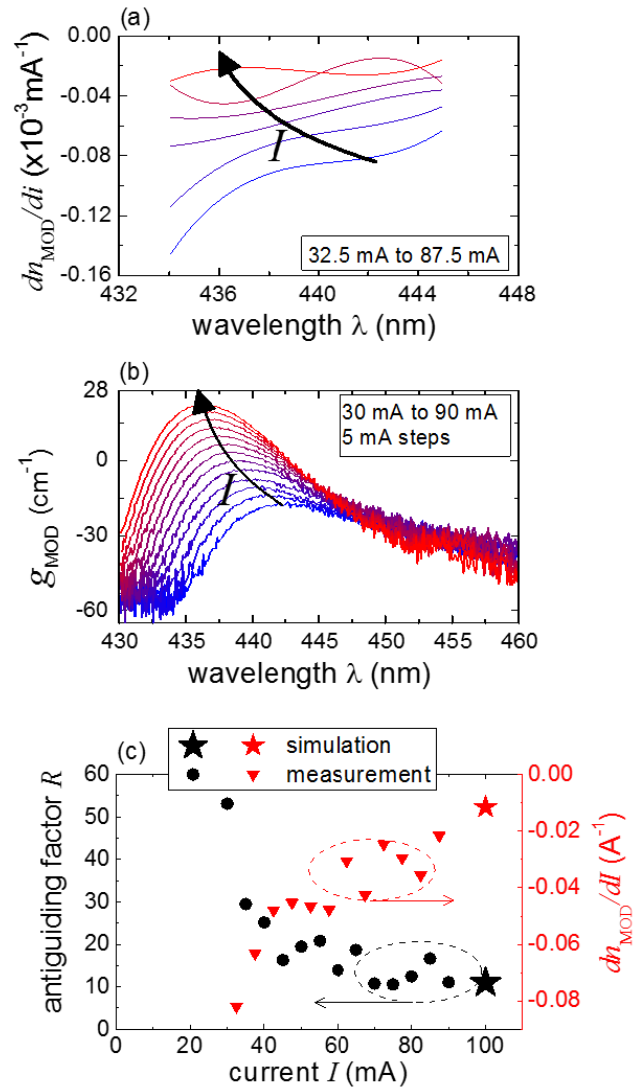

Fig. 3. (a) Modal gain spectra obtained by the Hakki-Paoli method. (b) Differential carrier-induced modal index spectra; the curves are polynomial fits of the measured data points. For clarity, only six of the eleven measured curves are shown here. In both (a) and (b) the black arrows mark the shift of the peak modal gain by increasing current. (c) Differential carrier-induced modal index at peak gain (right axis, red triangles) and antiguiding factor (left axis, black circles) as function of the current. The stars mark the values assumed in the simulations to reproduce the measurement data.

## III. Simulation Results and Discussion

2D electro-optical simulations were performed with the software LASTIP by Crosslight Software Inc. [16]. The software solves self-consistently the Schrödinger equation, the Poisson equation, and the equations for charge transport and optical waveguiding; current spreading is taken into account as well. The hole mobility in the p-type layers was set to 10 cm$^2$ V$^{-1}$ s$^{-1}$. Only half of the ridge was simulated due to the symmetry of the device. The lateral width of the simulated area was at least 6 µm and perfectly matched layers (PMLs) were inserted at the boundaries. The imaginary part of the relative permittivity in the PMLs was 0.1, while their thickness was set to 6 µm. The thickness of the PMLs was chosen large enough to avoid unwanted reflections at the domain boundaries. The structure included the Pd contact metal, the $Si_3N_4$ insulating layer and the Ti contact pad. No substrate was taken into account. Only the real parts of the refractive indexes of the epitaxial layers were used. Optical absorption was taken into account by assuming a modal loss parameter $\alpha_l$ = 35 cm$^{-1}$, as determined by the Hakki-Paoli method [compare Fig. 3(a)]. The carrier-density-dependent part of the refractive index in the quantum wells was calculated by

$$\Delta n_{r,N} = \frac{1}{2}\frac{dn_r}{dN}(n + p - n_0 - p_0), \quad (3)$$

where $n$ and $p$ are the electron and hole densities under biased conditions, and $n_0$ and $p_0$ are the electron and hole densities at thermal equilibrium. Note that under lasing $n \approx N + n_0$ and $p \approx N + p_0$ hold in the quantum wells. The effect of the polarization charges at the interface boundaries was taken into account by using the model by Fiorentini *et al.* [17] and assuming 50 % compensation by charged defects. The Poole-Frenkel field ionization of acceptors was also included. Considering the short current pulses used in the measurements, heating effects were neglected.

First, the threshold current of the deep-ridge device was adjusted to the experimental value by varying the parameter $C$ of the Auger recombination rate

$$R_{Aug} = C(n + p)(np - n_0 p_0). \quad (4)$$

Assuming $C$ = 1.7 × 10$^{-30}$ cm$^{-6}$/s, $I_{th}$ = 100 mA was obtained, which corresponds nearly to the measured average threshold current of the deep-ridge laser diodes.

Next, the threshold current $I_{th}$ of the shallow-ridge laser diode was calculated using the same parameters. In contrast to the deep-ridge, $I_{th}$ of the shallow-ridge strongly depends on the antiguiding parameter. Using $dn_r/dN$ = − 58 × 10$^{-22}$ cm$^3$, $I_{th}$ = 270 mA was obtained for the shallow-ridge device, which roughly agrees with the average experimental value. On the contrary, neglecting the antiguiding effect, $I_{th}$ for the shallow ridge was only slightly larger than for deep-ridge. The calculated *L–I* curves are shown in Fig. 4 (a).

A comparable increase of the threshold current could be obtained by a hypothetical increase of the hole mobility in the p-doped layers from 10 to 500 cm$^2$/Vs, resulting in an increased current spreading effect. However, as discussed in Ref. [18], no experimental investigation could find any evidence of a significant difference in current spreading between the shallow-ridge and deep-ridge devices. Additionally, the side lobes visible in the measured far-field patterns were not obtained in the simulation. Figures 4 (b) and 4 (c) show the lateral far-field patterns for both devices calculated using the antiguiding coefficient reported above. In the shallow-ridge case, side lobes are obtained which are very similar to the ones observed in experiment [compare Fig.

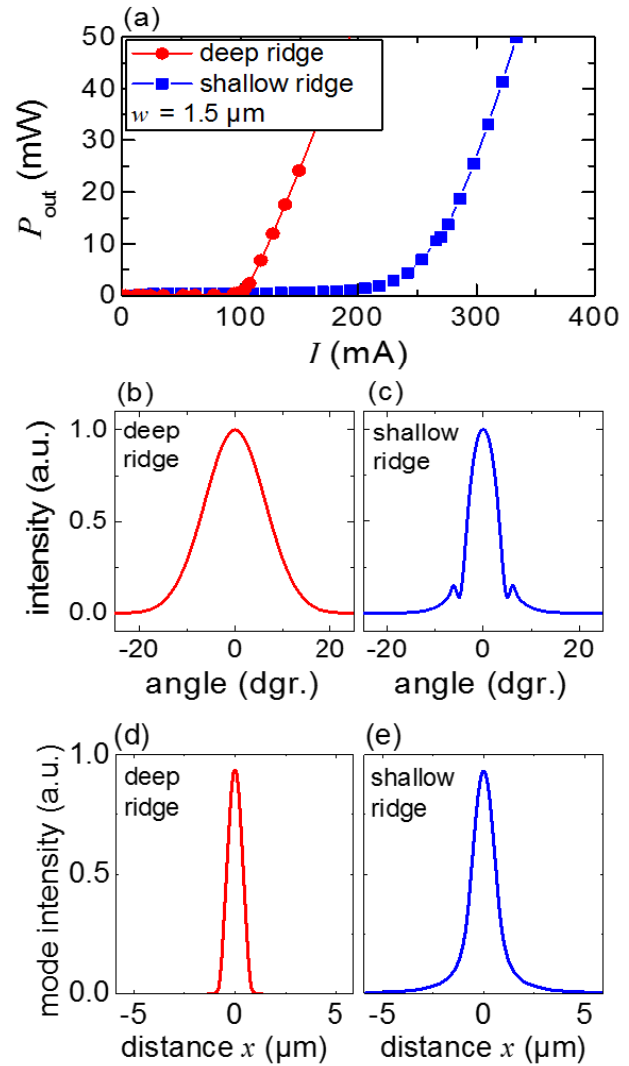

Fig. 4. Calculated *L−I* characteristics (a) and lateral far-field patterns (b) and (c) of the deep-ridge and shallow-ridge laser diodes assuming an antiguiding coefficient $dn_r/dN = -58 \times 10^{-22}$ cm$^{-3}$. In (d) and (e) the mode intensity in the top QW is plotted as a function of the lateral distance from the ridge center *x*.

2 (c)]. In Fig.4 (d) and (e) the calculated lateral mode intensity distribution is shown: note that in the shallow-ridge case the mode is broader and has much longer tails, extending almost over the whole simulation domain, due to the weaker guiding. These tails are the origin of the side lobes visible in the far-field pattern.

The effect of the large antiguiding coefficient on the lateral mode confinement can be illustrated by considering the carrier-induced effective index change of the shallow-ridge laser diode as a function of $x$ after integration in transverse direction $y$, i.e.

$$\Delta n_{\text{eff},N}(x) = \frac{\int_{-\infty}^{+\infty} |E(0,y)|^2 \Delta n_{r,N}(x,y) dy}{\int_{-\infty}^{+\infty} |E(0,y)|^2 dy}. \quad (5)$$

The obtained effective index profile is shown in Fig. 5. In the center of the ridge ($x$ = 0) the absolute value of the carrier-induced change of the effective index ($\Delta n_{\text{eff},N}(x)$ = –0.0058) is larger than the built-in index step $\Delta n_{\text{eff,shallow}}$ = 0.004. It can be therefore concluded that the lateral leakage of the optical mode due to index-antiguiding is responsible for the increased threshold current and the appearance of lateral side-lobes in the far-field pattern of shallow ridge lasers. Note that, not only the calculated carrier-induced change of the effective index of the deep-ridge laser ($\Delta n_{\text{eff},N}(x)$ = –0.0043) is smaller than the corresponding value of the shallow-ridge laser due to the much lower carrier density at threshold, but it is also much smaller than the built-in index step $\Delta n_{\text{eff,deep}}$ = 0.026. It is interesting to note that the carrier-induced index depression in Fig. 5 extends over several micrometers beyond the ridge. This effect is due to the lateral spreading of the charge carriers, which is enhanced by increasing the drive current and the carrier density in the MQW.

In order to judge how far the value assumed for $dn_r/dN$ in the simulations is realistic, the corresponding antiguiding factor $R$ of the simulated deep-ridge laser diode was calculated. It is important to note that the assumption of a constant $dn_r/dN$ in the simulation is an approximation for real devices for which $dn_r/dN$ should change with the current. Therefore, the value of the antiguiding factor is only derived from the simulations at threshold. The values of $\lambda$, $dn_{\text{MOD}}/dI$, and $dg_{\text{MOD}}/dI$ were obtained as a function of the current from the simulations, and substituted in (2). At $I_{\text{th}}$ = 100 mA, a value $R$ = 11.0 was obtained. Considering the approximations, this number is in excellent agreement with the experimentally determined value of $R_{\text{exp}} \approx 10$. The values of $dn_{\text{MOD}}/dI$ and $R$ derived from the simulations are indicated by stars in Fig. 3(c).

## IV. Conclusions

InGaN MQW RW laser diodes with two different ridge etch depths were investigated. A relatively small reduction in the etch depth (70 nm) resulted in a large increase in the threshold current density by a factor 2.5 to 3 as well as the appearance of side lobes in the lateral far-field patterns. Using the Hakki-Paoli method an antiguiding factor $R_{\text{exp}} \approx 10$ was determined experimentally, which should result in strong antiguiding effects. Self-consistent simulations show that, assuming a carrier-induced index change which is in agreement with the measured antiguiding factor, both the large threshold and the side-lobes in the far-field pattern can be reproduced.

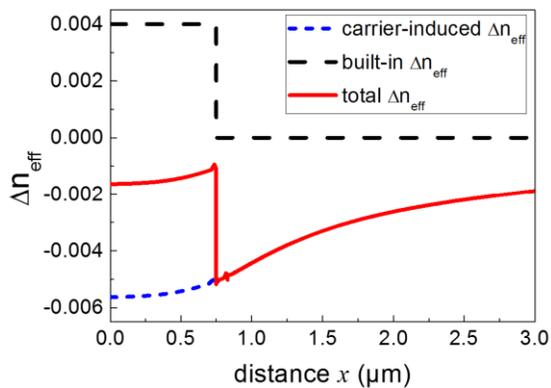

Fig. 5. Calculated effective index profile for the shallow-ridge laser diode above the lasing threshold. The total $\Delta n_{\text{eff}}$ is calculated by adding the built-in index step, due to the ridge waveguide, to the carrier-induced index change.

## Biographies

**Luca Redaelli** received his M.Sc. in electronic engineering from the Politecnico di Milano, Italy, in 2009 and his Ph.D. in electrical engineering from the the Technische Universität Berlin (TUB), Germany, in 2013. During his Ph.D. he worked on the design and fabrication of nitride-based laser diodes at the Ferdinand-Braun-Institut, Leibniz-Institut für Höchstfrequenztechnik in Berlin. He is currently working as a researcher for the French Alternative Energy and Atomic Energy Commission (CEA) in Grenoble, France, and has been awarded the Maria Sklodowska-Curie Fellowship by the European Commission for the years 2015/2016. His research interests span the fields of optoelectronics, micro and nanotechnology, with a particular focus on GaN-based devices such as lasers, LEDs, and solar cells. Further topics of interest are superconducting-nanowire single-photon detectors and quantum information technology.

**Hans Wenzel** received his Diploma and Doctoral degrees in physics from Humboldt University in Berlin, Germany, in 1986 and 1991, respectively. His thesis dealt with the electro-optical modeling of semiconductor lasers. From 1991 to 1994, he was involved in a research project on the simulation of distributed feedback lasers. In 1994, he joined the Ferdinand-Braun-Institut, Leibniz-Institut für Höchstfrequenztechnik in Berlin, where he is engaged in the development of high-brightness semiconductor lasers. He authored or coauthored more than 250 journal papers and conference contributions. His main research interests include the analysis, modeling, and simulation of optoelectronic devices.

**Joachim Piprek** received his Ph.D. in theoretical physics from Humboldt University in Berlin, Germany. For more than two decades, he worked in industry and academia on design, simulation, and analysis of various semiconductor devices used in optoelectronics. Dr. Piprek has taught graduate courses at universities in Germany, Sweden, and in the United States. He was invited guest editor for several journal issues on optoelectronic device simulation and currently serves as an executive editor of Optical and Quantum Electronics as well as an associate editor for the Journal of Computational Electronics. Dr. Piprek is founder and co-chair of the annual conference on Numerical Simulation of Optoelectronic Devices and he has also (co-)chaired several SPIE conferences. He has published 3 books, 6 book chapters, 4 patents, and more than 200 papers with more than 5000 citations.

**Thomas Weig** received his B.Sc. degree in physics from the Universität Augsburg, Germany in 2009 and his M.Sc. degree in physics from the Universität Stuttgart, Germany in 2011. Since then he works with the Fraunhofer Institute for Applied Solid State Physics (IAF), in Freiburg, Germany pursuing the PhD degree. His research interests include the development and characterization of (In,Al)GaN-based laser diodes emitting in the blue and ultraviolet region and, specifically, the ultra-short pulse generation in these devices.

**Gerrit Lükens** received his B.Sc. and M.Sc. degrees (with honors) in electrical engineering from Aachen University (RWTH), Aachen, Germany, in 2011 and 2013 respectively, where he is currently pursuing the Ph.D. degree. From 2012 to 2013 he held his internship at Fraunhofer IAF. Since 2014, he has been research assistant with RWTH Aachen University. His research focuses on GaN-based microelectronics and process technology.

**Martin Martens** received his diploma degree in physics (Dipl.-Phy) from the Technische Universität Berlin, Germany, in 2010, where he is currently pursuing the PhD degree. His research interests include the optical and electrical characterization of III-nitride light emitters, and in particular of (In,Al)GaN-based laser diodes emitting in the blue and ultraviolet regions of the spectrum.

**Sven Einfeldt** received his Diploma in Physics at the University of Jena (Germany) in 1990 and his doctor's degree (PhD) at the University of Würzburg (Germany) in 1995 for his work on the epitaxy of II-VI compounds. Until 2004 he worked as a post-doc at the University of Bremen (Germany) and at North Carolina State University (U.S.A.) focusing on the epitaxial growth and characterization of group III-nitrides and the fabrication of corresponding laser-diodes. He received the postdoctoral lecture qualification (Habilitation) in 2002. Since 2004 he is with the Ferdinand-Braun-Institut, Leibniz-Institut für Höchstfrequenztechnik (FBH), in Berlin (Germany). Currently, his main interests are brilliant and high-power short-wavelength laser diodes and deep ultraviolet light emitting diodes.

**Ulrich T. Schwarz** received his Ph.D. degree and habilitation in physical science from the University of Regensburg, Germany, in 1997 and 2004, respectively. Since 2009 he is full professor (W2) at the Institute for Microsystems Engineering (IMTEK) at Freiburg University. At the same time he is group leader at the Fraunhofer Institute for Applied Solid State Physics (IAF) in Freiburg. His research is in the field of optoelectronic devices based on group-III-nitrides, i.e. light emitting diodes and laser diodes in the violet to green spectral region. Another focus is on singular optics, in particular polarization singularities. Awarded by the Alexander von Humboldt foundation with a Feodor Lynen scholarship he spent two postdoctorial years (1997–1999) at Cornell University, Ithaca, NY, with research on intrinsically localized modes. In 2001 he joined the group of Prof. R. Grober at Yale University, New Haven, CT. In 2006/2007, he visited Kyoto University, Kyoto, Japan, with an invited fellowship (long-term) awarded by the Japanese Society for the Promotion of Science. He is senior member of the Optical Society of America (OSA), and member of the American Physical Society (APS), International Society for Optical Engineering (SPIE), and the German Physical Society (DPG).

**Michael Kneissl** (M'96, SM'03) received his diploma and Ph.D. degree in physics from Friedrich-Alexander-University, Erlangen, Germany, in 1993 and 1996, respectively. He joined Xerox PARC in 1996 as Research Associate, became Member of the Research Staff in 1997 and was promoted to Principal Scientist in 2004. Since 2005 he is a Professor and Chair of "Experimental Nanophysics and Photonics" at the Technische Universität (TU) Berlin, Germany. He holds a joint appointment at the Ferdinand-Braun-Institut, Leibniz-Institut für Höchstfrequenztechnik (FBH) in Berlin, where he heads the GaN-Optoelectronics Business Area. Since 2010 he also is Executive Director of the Institute of Solid State Physics at TU Berlin. His research interests include group III-nitride semiconductors and nanostructures, metal organic vapor phase epitaxy of wide bandgap materials as well as novel optoelectronic devices, including UV LEDs and laser diodes. He has co-authored over 250 publications, four book chapters, and holds more than 50 patents.